\documentclass[aps,prb,groupedaddress]{revtex4}
\usepackage{epsfig}
\usepackage[american]{babel}
\usepackage{amsmath}
\usepackage[dvips]{color}

\usepackage[latin2]{inputenc}
\usepackage[T1]{fontenc}
\usepackage[american]{babel}
\usepackage{color}
\usepackage{graphicx}
\usepackage{graphics}

\newcommand{\Sc}{Schr\"{o}dinger}

\begin{document}

\title{Relevant coherent states for quantum description of adiabatic polarons}

\author{O. S. Bari\v si\' c}


\affiliation{Institute of Physics, Bijeni\v cka c. 46, HR-10000 Zagreb, Croatia}

\begin{abstract}

A new numerical method is proposed for determining the low-frequency dynamics of the charge carrier coupled to the deformable quantum lattice. As an example, the polaron band structure is calculated for the one-dimensional Holstein model. The adiabatic limit on the lattice, which cannot be reached by other approaches, is investigated. In particular, an accurate description is obtained of the crossover between quantum small adiabatic polarons, pinned by the lattice, and large adiabatic polarons, moving along the continuum as classical particles. It is shown how the adiabatic contributions to the polaron dispersion, involving spatial correlations over multiple lattice sites, can be treated easily in terms of coherent states.

\end{abstract}

\maketitle

The motion of charge carriers, coupled to the phonon field on a lattice, has been an object of intensive investigations for many years. In this context, different types of quasi-particles are studied, such as polarons, bipolarons, topological solitons, to name a few. In particular, polarons were first introduced as states of broken translational symmetry, describing a localized electron trapped by the self-induced, static lattice deformation \cite{Landau}. By taking the lattice kinetic energy into account, the translational symmetry is restored: the deformation moves along the lattice followed almost instantaneously (adiabatically) by the electron, provided that it is fast enough. For large adiabatic polarons, which behave as free particles, the effective mass can be obtained in the continuous approximation, treating the lattice classically \cite{Pekar}. In the opposite (small polaron) limit, the polaron is strongly pinned by the discreteness of the lattice. For this latter case it is well-known that the quantum treatment is necessary to describe the dispersion properly \cite{Holstein}. Although the translationally invariant diagrammatic perturbation theory \cite{OSSB} or similar methods provide in principle the unifying approach to the crossover between large and small (adiabatic) polarons, the corresponding calculations are intricate and have not yet been carried out.

This long-standing problem can be simply solved by the relevant coherent states method (RCSM) developed here. As an illustrative example, the RCSM is applied to the 1D Holstein polaron, involving an electron coupled locally to the lattice displacement. The Holstein Hamiltonian is given by \cite{Holstein}

\begin{equation}
\hat H=-t\sum_nc_n^\dagger(c_{n+1}+c_{n-1})+
\omega_0\sum_nb_n^\dagger b_n-g\sum_nc_n^\dagger c_n(b_n^\dagger+b_n)\;,\label{HolHam}
\end{equation}

\noindent with $c_j^{\dagger}$ ($c_j$) and $b^{\dagger}_j$ ($b_j$) the creation (annihilation) operators for the electron and phonon at the site $j$, respectively. $t$ is the nearest-neighbor electron hopping integral, $\omega_0$ is the energy of the dispersionless optical phonon branch, and $g$ is the electron-phonon coupling constant. There have been extensive studies \cite{Feinberg} of the polaron states for the Hamiltonian~(\ref{HolHam}). Accurate results were obtained for values of the adiabatic ratio $t/\omega_0\lesssim5$, which are far too small to address the formation of the large adiabatic polaron. On the other hand, the RCSM does not suffer from such restriction. With modest computational effort, it can be applied for any value of $t/\omega_0$, including the $t/\omega_0\gg1$ limit. In this way, the crossover between small and large adiabatic polarons can be investigated.

The RCSM is basically a variational procedure performed in two steps. In the first one, a finite set of localized polaron wave functions $|\varphi_j^s\rangle$ is selected, capable of describing the polaron adiabatic motion within the unit cell surrounding the site $j$. For the site $j$, the localized polaron wave function is expressed as their linear combination $|\varphi_j\rangle=\sum_sa_s|\varphi_j^s\rangle$, with variational coefficients $a_s$. In the second step, the translational invariance of the problem is restored by writing the polaron wave function in the translationally invariant form,

\begin{equation}
|\Psi_K\rangle=1/\sqrt N\;\sum_j e^{iKj}
\sum_s a_s|\varphi_j^s\rangle
=\sum_s a_s|\Psi_K^s\rangle\;,\label{Psiglobal}
\end{equation}

\noindent with $K$ the polaron momentum. The set of wave functions $|\Psi_K^s\rangle$ is linearly independent provided that 
$|\varphi_j^s\rangle\neq|\varphi_{j'}^{s'}\rangle$
for all $j\neq j'$ and $s\neq s'$. The minimization of the variational energy with respect to the complex coefficients $a_s^*$ yields the generalized eigenvalue problem, 

\begin{equation}
\sum_{s'}\langle\Psi_K^s|\hat H|\Psi_K^{s'}\rangle\;a_{s'}=
\sum_{s'}\langle\Psi_K^s|\Psi_K^{s'}\rangle\;a_{s'}\;,\label{GEP}
\end{equation}

\noindent the solutions of which are the lowest and excited polaron states for a given momentum $K$.

For a given set of parameters the RCSM trial wave function (\ref{Psiglobal}) is chosen according to the properties of the effective potential that characterizes the adiabatic dynamics of the polaron. This makes the RCSM essentially different from the other variational approaches in the literature. Each localized polaron wave function $|\varphi_j^s\rangle$ in eq.~(\ref{Psiglobal}) is formed as a product of the electron and the lattice part,

\begin{equation}
|\varphi_j^s\rangle=
\left[\sum_n\eta_n(\vec x^s)\;c_{j+n}^\dagger\right]\;
\hat S_j(\vec x^s+i\vec p^s)\;|0\rangle\;.\label{VFMZKS}
\end{equation}
 
\noindent The wave function (\ref{VFMZKS}) is parameterized by the $N$-dimensional complex vector $\vec x^s+i\vec p^s=\{x_n^s,p_n^s\}$, where $N$ is the number of lattice sites ($N\rightarrow\infty$). The lattice part in eq.~(\ref{VFMZKS}) is given in terms of the coherent states, i.e., of the displaced harmonic oscillators,

\[\hat S_j(\vec x^s +i\vec p^s)|0\rangle\equiv\prod_n
\exp{\left[(x_n^s +ip_n^s)
(b_{j+n}^\dagger-b_{j+n})\right]}|0\rangle\;.\]

\noindent Here, the vector component $x_n^s$ defines the mean lattice displacement at the site $j+n$, whereas $p_n^s$ defines the mean lattice momentum. In eq.~(\ref{VFMZKS}), the electron state $\eta_n(\vec x^s)$ is given by the electron ground state corresponding to the mean lattice deformation $\vec x^s$. In this respect, the method of calculating $\eta_n(\vec x^s)$ in eq. (\ref{VFMZKS}) follows the same steps as the adiabatic approximation~\cite{Kalosakas}. In particular, the electron wave function is obtained as the ground state $\nu=0$ solution of the equation

\begin{equation}
\varepsilon^{(\nu)}(\vec x^s)\;\eta_n^{(\nu)}=
-t\;\left(\eta_{n+1}^{(\nu)}+\eta_{n-1}^{(\nu)}\right)
-2g\;x_n^s\;\eta_n^{(\nu)}\;. 
\label{ElSP}
\end{equation}

\noindent With $\eta_n\equiv\eta_n^{(0)}$ in eq.~(\ref{VFMZKS}), $\sum_n|\eta_n|^2=1$, the expectation value of the Hamiltonian (\ref{HolHam}) is obtained for the localized polaron wave function (\ref{VFMZKS}) as

\begin{equation}
\langle\varphi_j^s|\hat H|\varphi_j^s\rangle =\omega_0|\vec p^s|^2+\omega_0|\vec x^s |^2+\varepsilon(\vec x^s)\;,\label{UFunc}
\end{equation}

\noindent where $\varepsilon(\vec x^s)\equiv\varepsilon^{(0)}(\vec x^s)$ is the electron ground state energy of eq.~(\ref{ElSP}). The last two terms in eq. (\ref{UFunc}) define the adiabatic potential $U_{AD}(\vec x)$, i.e., the potential energy of the lattice in the adiabatic approximation. This simple relation between the wave function (\ref{VFMZKS}) and the adiabatic potential provides a well-controlled and efficient way of treating the adiabatic correlations.

Although $\vec x^s$ can represent any lattice deformation in eq.~(\ref{VFMZKS}), it is important to realize that the set of wave functions $|\varphi_j^s\rangle$ in eq.~(\ref{Psiglobal}) is meant to capture the polaron dynamics around the site $j$, whereas the delocalization of the polaron over the entire lattice is accounted for by the translationally invariant form of the trial wave function~(\ref{Psiglobal}). Furthermore, it should be stressed that the trial wave function~(\ref{Psiglobal}) includes nonadiabatic processes as well. Namely, each localized polaron wave function $|\varphi_j^s\rangle$ in eq.~(\ref{VFMZKS}) fulfills the adiabatic approximation only to the mean-field level. That is, in eq.~(\ref{VFMZKS}) the localized electron wave function $\eta_n(\vec x^s)$ is determined through eq.~(\ref{ElSP}) by the mean lattice deformation $\vec x^s$. On the other hand, the lattice quantum fluctuations around $\vec x^s$ associated with coherent states, in eq.~(\ref{VFMZKS}) are not followed adiabatically by the electron. In particular, one can easily verify that the RCSM reproduces the well-known small nonadiabatic polaron~\cite{Holstein}. For $\lambda=\varepsilon_p/t\gg1$, the polaron lattice deformation localizes adiabatically to a single lattice site, $x_n^s\approx g/\omega_0\;\delta_{n,0}$, where $\delta_{n,0}$ is the Kronecker delta. By considering solely this lattice configuration corresponding to the minimum of the adiabatic potential, eq.~(\ref{ElSP}) gives $\eta_n^{(0)}\approx\delta_{n,0}$, which yields, through eq.~(\ref{Psiglobal}), the polaron wave function~as 

\[|\Psi_K\rangle\approx
1/\sqrt N\;\sum_j e^{iKj}\;
c_j^\dagger\;e^{g/\omega_0\;(b_j^\dagger -b_j)}|0\rangle \;.\]

\noindent The corresponding expectation value of the energy~(\ref{GEP}) is equal to the small nonadiabatic polaron energy $-\varepsilon_p-\varepsilon_K\;e^{-\varepsilon_p/\omega_0}$, with $\varepsilon_K$ the free electron energy. Such dispersion of the polaron is purely nonadiabatic: during the hopping to the nearest-neighbor sites the electron and the lattice deformation detach from each other.

When the adiabatic contributions to the polaron dispersion, described by a joint motion of the electron and the lattice deformation, are dominant the situation is fundamentally different. In particular, the adiabatic motion of the polaron between lattice sites is characterized by two types of stationary points of $U_{AD}(\vec x)$, the minima $\vec x_M$ and the saddle points with the lowest energy $\vec x_{PN}$ (i.e., two types of lattice deformations). The first correspond to polarons centered at the lattice site, whereas the second correspond to polarons centered half-way between the lattice sites. Their difference in energy defines the Peierls-Nabarro (PN) barrier~$\Delta_{PN}$,

\[\Delta_{PN}=U_{AD}(\vec x_{PN})-U_{AD}(\vec x_M)\;,\]

\noindent which is the minimal energy barrier that must to be overcome in order to move the polaron classically and adiabatically from one lattice site to another.

The adiabatic motion of the large polaron within one unit cell involves changes of the lattice deformation over many lattice sites. This difficulty of dealing with a large number of lattice coordinates can be avoided by introducing a smaller set of new coordinates that are most relevant for the polaron dispersion. One of the possible solution is to work with the moving set of coordinates $\xi$ and $Q_\alpha(\xi)$, where $\xi$ is the centroid coordinate determining the position of the polaron along the minimal energy path of its motion, while $Q_\alpha(\xi)$ are the coordinates of the normal modes $\alpha$ that are moving with the polaron \cite{Miller,HolsteinTurkevich}. The efficiency of this approach is derived from the fact that the effective coupling of $\xi$ with $Q_\alpha(\xi)$ is significant only for a few modes $\alpha$.

While in the RCSM context, the moving set of coordinates can be used quite generally for various electron-phonon models. An alternative choice of coordinates is considered here for the Holstein model, based on the specific properties of the adiabatic potential. The homogenous lattice deformation in eq. (\ref{HolHam}) couples only to the total electron density. Consequently, the harmonic dynamics of the homogenous $q=0$ lattice mode can be separated out, and one has to consider explicitly only the lattice deformations satisfying the sum rule $\sum_nx_n=g/\omega_0$ \cite{Barisic2}. For these physically interesting deformations $\vec x=g/\omega_0\;\hat x$, where $\hat x$ denotes the $N$-dimensional unit vector, it follows from eq.~(\ref{ElSP}) that $U_{AD}(\vec x)$ exhibits a simple scaling behavior, to our knowledge not noted before, 

\begin{equation}
U_{AD}(\vec x)=\omega_0|\vec x|^2+\varepsilon(\vec x)=
\varepsilon_p\;U_{AD}(\hat x, \lambda)\;,\label{Scaling}
\end{equation}

\noindent with $\varepsilon_p=g^2/\omega_0$ and $\lambda=\varepsilon_p/t$ determining respectively the amplitude and the shape of the adiabatic potential. One immediately recognizes that $\lambda$ governs the unit vectors corresponding to stationary points, given by $\vec\nabla U_{AD}(\vec x)=0\Rightarrow x_n=g/\omega_0\;|\eta_n|^2$. These points can be found \cite{Proville} from the discrete nonlinear \Sc\ equation (\ref{ElSP}). Furthermore, $\lambda$ governs the harmonic adiabatic dynamics around the minima of $U_{AD}(\vec x)$. The expansion of $U_{AD}(\vec x)$ in powers of displacements $\vec\delta$ from the equilibrium deformation $\vec x_M$, $\vec \delta=\vec x-\vec x_M$, is given by

\[U_{AD}(\vec x)=U_{AD}(\vec x_M)
+\omega_0\sum_{n,m}\Pi_{n,m}\;\delta_n \delta_m +U_{AD}^{AH}\;.\]

\noindent The second term determines the harmonic fluctuations, while the adiabatic anharmonic dynamics is described by the higher order contributions $U_{AD}^{AH}$. The Hessian matrix $\Pi_{n,m}$ depends on $\lambda$, and it can be expressed by means of the electron basis (\ref{ElSP}) corresponding to $\vec x_M$ \cite{Kalosakas,HolsteinTurkevich},

\[\Pi_{n,m}=\delta_{n,m}
-4\;\varepsilon_p\;\eta_n^{(0)}\eta_m^{(0)}
\sum_{\nu\neq0}\frac{\eta_n^{(\nu)}\eta_m^{(\nu)}}
{\varepsilon^{(\nu)}-\varepsilon^{(0)}}\;,\] 

\noindent with $\delta_{n,m}$ the Kronecker delta. Up to the second order in $\vec\delta $, the lattice vibrations decouple in terms of normal modes $\alpha$, which are characterized by normal coordinates $\delta_\alpha$ and eigenenergies $\omega_\alpha$. Notice that both, the unit displacement $\hat\delta_\alpha$ along the normal coordinate $\delta_\alpha$ and the ratio $\omega_\alpha/\omega_0$, with $\omega_0$ the bare phonon energy, are determined by $\lambda$.  

\begin{figure}[tpb]

\begin{center}{\scalebox{0.3}{\includegraphics{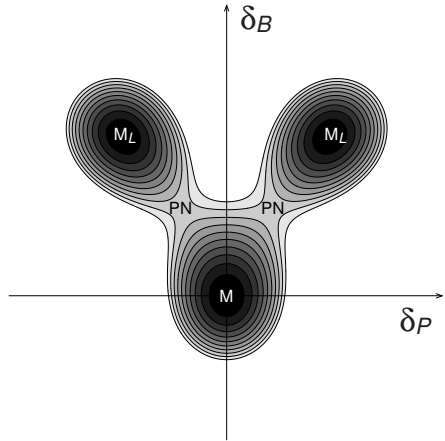}}}\end{center}
\caption{The adiabatic potential for the small polaron case ($\lambda=2$), as a function of coordinates $\delta_P$ and $\delta_B$, belonging to the pinning and breather normal modes. The absolute minimum, the two saddle points and the two local minima are denoted by $M$, $PN$, and $M_L$, respectively.\label{fig01}}

\end{figure}

As shown here, the use of normal coordinates $\delta_\alpha$ provides a simple and efficient way of keeping track of the lattice configurations that are relevant for the motion of the Holstein polaron associated with one unit cell. Namely, by considering the displacement from the equilibrium configuration to the nearest saddle point in terms of normal coordinates,

\[\vec\delta_{PN}=\vec x_{PN}-\vec x_M\sim g/\omega_0\;
\sum_\alpha\cos{\gamma_\alpha}\;\hat\delta_\alpha\;,\]

\noindent one finds that $\vec\delta_{PN}$ lies almost entirely in the plane of the two lowest normal modes. That is, for all values of parameter $\lambda$, which determine the arguments $\gamma_\alpha$, one obtains $\cos^2{\gamma_P}+\cos^2{\gamma_B}>0.99$. Here, $P$ (pinning) and $B$ (breather) are used to denote the two lowest, odd and even, normal modes. A better insight into the properties of the adiabatic potential can be obtained from fig.~\ref{fig01}, which shows the adiabatic potential for $\lambda=2$ as a function of displacements along the normal coordinates $\delta_P$ and $\delta_B$. The absolute minimum of the adiabatic potential is at the origin, while the saddle points $PN$ correspond to the top of the PN barrier. It is satisfactory that the two local minima $M_L$ are given by lattice configurations which are close to the equilibrium deformation of the polaron centered at the nearest-neighbor sites.

The above findings related to the properties of the adiabatic potential imply that in order to describe the adiabatic polaron motion properly it is sufficient to take into account the adiabatic correlations related to the pinning and breather modes in eq.~(\ref{VFMZKS}). Such a procedure breaks the translational symmetry of the adiabatic potential (\ref{Scaling}), 

\begin{equation}
U_{AD}(\vec x)\rightarrow U(\vec x_M)+U(\delta_P,\delta_B)+\omega_0\sum_{\alpha\neq P,B}\delta_\alpha^2\;.\label{LocAprox}
\end{equation}

\noindent $U(\delta_P,\delta_B)$, shown in fig.~\ref{fig01}, describes the part of the adiabatic potential $U_{AD}(\vec x)$ that is relevant for the polaron motion within one unit cell. The adiabatic correlations related to higher normal modes, $\alpha\neq P,B$, which weakly affect the polaron dispersion properties, are not of significant physical interest here. These nearly harmonic correlations result in the softening of the higher modes. If not ignored, this softening would be manifested by a set of eigenenergies $\omega_\alpha$ that replace $\omega_0$ in the third term of eq.~(\ref{LocAprox}), $\omega_\alpha\lesssim\omega_0$.

In the present case, having identified the two essential coordinates for the adiabatic motion of the Holstein polaron within one unit cell, $\delta_P$ and $\delta_B$, it is straightforward to proceed with the RCSM. One considers a set of wave functions (\ref{VFMZKS}) given by complex amplitudes $\vec x^s+i\vec p^s$ corresponding to the points on the grid lying in the plane of the two lowest normal modes,

\begin{equation}
\vec x^s =\vec x_M+n\;\Delta x\;\hat\delta_P+ m\;\Delta x\;\hat\delta_B\;,\;\;\;
\vec p^s =n'\;\Delta p\;\hat\delta_P+ m'\;\Delta p\;\hat\delta_B\;,\label{CSPN}
\end{equation}

\noindent with $n,m,n',m'$, integers, and $\Delta x$, $\Delta p$, grid spacings. The finite set of states (\ref{VFMZKS}), important for the low-frequency dynamics, can be chosen by considering only such $\vec x^s$ and $\vec p^s$ for which the energy (\ref{UFunc}) is below a certain value. For the polaron ground state, this value should be at least of the order of a few $\omega_0$, or of the order of the PN barrier if this barrier is large \cite{OSB103}. The grid spacing $\Delta x$ in eq.~(\ref{CSPN}), as it can be seen from eqs.~(\ref{Psiglobal}) and~(\ref{VFMZKS}), defines the number of different electron wave functions $\eta_n$ involved in calculations. The adiabatic correlations are described with better accuracy as this number is increased by decreasing $\Delta x$. However, when $\Delta x$ is too small, $\Delta x \ll1$, some functions (\ref{VFMZKS}) are nearly identical, and eq. (\ref{GEP}) cannot be solved numerically. Nevertheless, satisfactory results are obtained for $\Delta x,\Delta p\sim1$ already.

\begin{figure}[t]

\begin{center}\centering{\scalebox{0.5}{\includegraphics{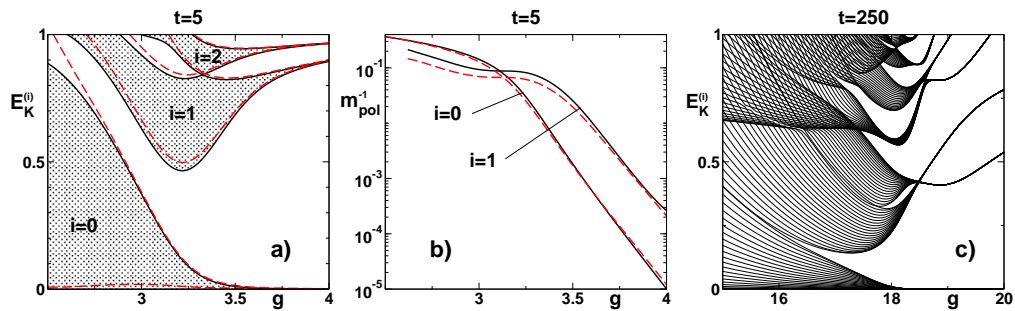}}}\end{center}
\caption{Panel a): the polaron band structure below the phonon threshold for $t=5$ resolved by the ETM (solid curves) and the RCSM (dashed curves). Panel b): ETM vs. RCSM polaron effective mass $m_{el}/m_{pol}^{(i)}$ for the two lowest $i=0,1$ bands ($t=5$). Panel c): the polaron band structure for $t=250$ in the crossover regime by means of 22 states with different momenta for each band, $K=0,\pi,n\times0.15$, $n\leq20$. $\omega_0$ is used as the unit of energy.\label{fig02}}

\end{figure}

In order to check the validity of the RCSM, a comparison with the results of the previously developed \cite{Trugman} exact translational method (ETM) is performed. For $t/\omega_0\lesssim5$ the ETM provides the polaron band structure below the phonon threshold with great accuracy \cite{Barisic2,Barisic3}. This makes it particularly suitable for the analysis of the RCSM, which exhibits the excited polaron bands as well. Figure~\ref{fig02}a shows the lowest $i=0$ and the two $i=1,2$ excited bands related to the pinning and the breather mode in terms of the band boundaries corresponding to $K=0$ and $K=\pi$ polaron states. All the curves are given with respect to the ETM ground state energy. The solid curves are the ETM, whereas the dashed are the RCSM results. The choice of parameters in fig.~\ref{fig02}a corresponds to the crossover regime between free and pinned polaron states. As discussed in detail in ref. \cite{Barisic3}, for $g/\omega_0\approx3.2$ there is a strong hybridization of the $i=1$ and $i=2$ excited bands. Such behavior may be explained from eq.~(\ref{LocAprox}) and fig.~\ref{fig01}: the two lowest normal modes are effectively coupled by the polaron motion. The ETM bands shown in fig.~\ref{fig02}a are reproduced very accurately by the RCSM. The differences become pronounced only close to the phonon threshold $E_K^{(i)}\approx\omega_0$, where long-range correlations mediated by weakly correlated phonon excitations become important due to the existence of the continuum of states above $\omega_0$ \cite{Barisic3}. Figure \ref{fig02}b shows the inverse effective mass of the polaron, $m_{el}/m_{pol}^{(i)}=\partial E_K^{(i)}/\partial\epsilon_K|_{K=0}$ for the same parameters as in fig. \ref{fig02}a. One finds that the agreement between the ETM (solid curves) and the RCSM (dashed curves) is excellent for the $i=0,1$ bands, even though $m_{pol}$ varies over four orders of magnitude.

The polaron crossover from pinned to free polaron states, shown in fig.~\ref{fig02}a for $t/\omega_0=5$, corresponds to an adiabatic-nonadiabatic crossover in the polaron dynamics. Namely, all the excited bands below the phonon threshold, related to the soft adiabatic phonon modes, at the weak-coupling side of fig.~\ref{fig02}a shift towards the phonon threshold, indicating that the adiabatic correlations become suppressed completely by the nonadiabatic contributions~\cite{Barisic3}. On the contrary, the polaron crossover shown in fig. \ref{fig02}c for $t/\omega_0=250$ is governed entirely by the adiabatic dynamics. Figure~\ref{fig02}c shows the lowest and the excited bands corresponding to the pinning and breather modes, shifted by the RCSM ground state energy. Each of the bands is given by the 22 states with different momenta, $K=0,\pi,n\times0.15$, $n\leq20$. On the left side of fig.~\ref{fig02}c the polarons are free (large), i.e., the PN is negligible. The polaron dispersion is approximately given by $E_K\sim K^2$ for the whole energy range shown in fig.~\ref{fig02}c.  For $g/\omega_0\approx15$ the energy of the $K=0$ state involving the excited breather mode is very close to the value derived in the continuous adiabatic approximation (large adiabatic polarons), $\omega_B/\omega_0=0.65$. On the right side of fig.~\ref{fig02}c the spectrum is characterized by a large PN barrier (small adiabatic polarons). That is, the bandwidths are exponentially small, while the position of the excited bands are given by the multiples of the pinning and breather mode energies, $m\omega_P+m'\omega_B$. The crossover between pinned and free adiabatic polarons takes place in the central part of fig.~\ref{fig02}c, where strong hybridization between excited bands occurs due to the pinning effects.

It is important to realize that the polarons at the left side of fig. \ref{fig02}c are fundamentally different from the polarons at the left side of fig.~\ref{fig02}a. While the latter are nonadiabatic and characterized by strong quantum effects, the effective mass of the former is predicted correctly by the classical soliton theory, i.e., as for the classical lattice deformation that moves along the lattice. The band structure in fig. \ref{fig02}c, exhibiting many polaron bands, has never been obtained before. It reveals how the dispersion of the adiabatic polarons changes its character from the quantum to the classical one, as the PN barrier decreases.

In conclusion, the newly proposed RCSM treats the low-frequency dynamics of charge carriers which interact with the quantum lattice. Its application to the 1D Holstein polaron problem provides the polaron band structure in the adiabatic limit $t\gg\omega_0$, not reachable by any other available method. In particular, the RCSM gives an accurate description of the crossover between the small and large adiabatic polarons, for which the nature of the polaron dispersion changes from quantum to classical one. For $t\sim\omega_0$ the accuracy of the RCSM is verified by comparison to the practically exact ETM results. The only differences of note are found only for weak couplings close to the phonon threshold when the long-range nonadiabatic correlations are important. As the weak-coupling regime can be described by the perturbation theory (PT) or ETM for any $t/\omega_0$, one finds that the RCSM together with the PT or ETM provides a solution of the Holstein polaron problem for the entire parameter space.


\begin{thebibliography}{99}
\bibitem{OSB103} In this work less than $10^3$ states are used.
\end{thebibliography}
\end{document}